\documentclass[letterpaper,twocolumn,10pt]{article}
\usepackage{usenix,epsfig,endnotes}

\usepackage{balance}
\usepackage{alltt}
\usepackage{amsmath}
\usepackage{balance}
\usepackage{booktabs}
\usepackage{fixltx2e}
\usepackage{graphicx}
\usepackage{boxedminipage}
\usepackage{hyperref}
\usepackage{nicefrac}
\usepackage{subfig}
\usepackage{setspace}
\usepackage{xspace}
\usepackage{multirow}
\usepackage{colortbl}
\usepackage{amsfonts} 
\usepackage{blindtext}
\usepackage{chngpage}
\usepackage{listings}
\usepackage{color}
\usepackage[dvipsnames]{xcolor}
\usepackage{mathtools}
\usepackage{amssymb}
\usepackage{pifont}
\usepackage[numbers,sort&compress,square]{natbib}

\usepackage{array}
\newcolumntype{x}[1]{>{\centering\arraybackslash\hspace{0pt}}p{#1}}

\RequirePackage{listings}

\lstdefinelanguage{Golang}%
  {morekeywords=[1]{package,import,func,type,struct,return,defer,panic,%
     recover,select,var,const,iota,},%
   morekeywords=[2]{string,uint,uint8,uint16,uint32,uint64,int,int8,int16,%
     int32,int64,bool,float32,float64,complex64,complex128,byte,rune,uintptr,%
     error,interface},%
   morekeywords=[3]{map,slice,make,new,nil,len,cap,copy,close,true,false,%
     delete,append,real,imag,complex,chan,},%
   morekeywords=[4]{for,break,continue,range,goto,switch,case,fallthrough,if,%
     else,default,},%
   morekeywords=[5]{Println,Printf,Error,Print,},%
   sensitive=true,%
   morecomment=[l]{//},%
   morecomment=[s]{/*}{*/},%
   morestring=[b]',%
   morestring=[b]",%
   morestring=[s]{`}{`},%
   }

\lstdefinelanguage{Rust}%
  {morekeywords={again,assert,break,const,do,drop,else,enum,%
      export,extern,fn,for,if,import,let,match,mod,new,%
      return,struct,trait,type,use,while},%
   sensitive,%
   morecomment=[s]{/*}{*/},%
   morecomment=[l]//,
   morestring=[b]",%
   morestring=[b]'%
  }[keywords,comments,strings,directives]%

\newcommand{\beforecaption}{\vspace{-.15cm}\begin{spacing}{0.85}}
\newcommand{\aftercaption}{\vspace{-.45cm}\end{spacing}}
\newcommand{\mycaption}[3]{\beforecaption\caption{\label{#1}{\textbf{#2}} \textit{\small{#3}}}\aftercaption}

\begin{document}

\twocolumn[
\begin{@twocolumnfalse}
\begin{center}
{\Large\bf Fearless Concurrency? Understanding Concurrent Programming Safety in Real-World Rust Software}
\end{center}
\vspace{-0.2in}
\begin{center}
Zeming Yu$\dagger$,
Linhai Song$\dagger$,
Yiying Zhang$\ast$ \\
$\dagger$ Pennsylvania State University, $\ast$ Purdue University 
\end{center}

\bigskip
\end{@twocolumnfalse}
]

\begin{abstract}
Rust is a popular programming language in building 
various low-level software in recent years. 
It aims to provide safe concurrency when implementing multi-threaded 
software through a suite of compiler checking rules.
Unfortunately, there is limited understanding of how 
the checking rules influence the safety of concurrent 
programming in Rust applications.

In this paper, we perform a preliminary study on Rust's concurrency
safety from two aspects:
concurrency usage and concurrency bugs. 
Our study can provide better understanding on Rust's concurrency
and can guide future researchers and practitioners in 
writing better, more reliable Rust software and in developing debugging 
and bug detection tools for Rust.

\end{abstract}

\section{Introduction}
\label{sec:intro}

Rust is a programming language designed to build efficient \textit{and} safe low-level software~\cite{rust-wiki, safe-concurrency-1,safe-concurrency-2,safe-memory-1,rust-performance-1}. 
Rust is targeted to achieve performance that is comparable to C, 
while avoiding many safety issues in C including both concurrency bugs and memory safety bugs.
Over the past few years, Rust has gained significant popularity,
especially in building low-level software~\cite{rust-love1,rust-love2,rust-love3}.
These Rust applications range from simple libraries~\cite{rand} 
and utilities~\cite{rust-binutil,rust-coreutils} 
to complex browsers~\cite{servo,quantum} 
and operating systems~\cite{stratis,tockos,redox}. 

A major design goal of Rust is to provide safe concurrency~\cite{safe-concurrency-1,safe-concurrency-2}.
To achieve the goal, Rust leverages several unique ownership rules, 
and imposes the rule checking during compilation. 
It is thus widely believed that Rust is much safer than many other languages.
As stated in Rust's official tutorial, ``concurrency is fearlesss'' in Rust.
Since Rust is widely used to implement low-level systems,
it is important to understand whether these compiler
rules can really reduce concurrency bugs and provide safer concurrency. 

Unfortunately, there is only limited prior work in understanding Rust's 
concurrency. Many key questions are left open. 
For example, how does Rust's compiler rules influence its programmability? 
Will developers have to use unsafe code to bypass compiler checkings?
What are the common mistakes made by developers when 
they develop multithreaded Rust programs? 
Will bypassing Rust's compiler checkings introduce concurrency bugs? 
Will there be no concurrency bugs in safe code? 
The lack of knowledge in Rust's concurrency would severely impair
Rust software development and make 
it hard to create Rust development and debugging tools.

In this paper, we conduct the first empirical study on Rust's concurrency. 
Our study focuses on two main aspects, 
concurrency usage (Section~\ref{sec:usage}) and concurrency bugs (Section~\ref{sec:bug}). 
We conduct our study on three popular open-source Rust applications: Servo~\cite{servo}, 
a web browser, TiKV~\cite{tikv}, a key-value storage system, 
and Rand~\cite{rand}, a random number generation library. 

Although our study is just preliminary, we are already able to identify interesting new
findings and insights in Rust concurrent programming. 
For example, we find that there are still data race bugs in safe Rust code
and Rust programmers should not rely solely on Rust's compiler checkings 
and be ``fearless'' in writing concurrent Rust programs.
Another interesting finding is Rust's compiler checkings do restrict its programmability 
and change developers' usage habits 
of concurrency primitives. 

Overall, we make four observations in our study of concurrent programming in Rust
and infer four implications that can be useful for future Rust programmers and language designers.
For example, future research in building Rust bug detectors should 
focus on unsafe code and atomic operations in safe code. 
Our findings and implications improve the understanding of Rust's 
concurrency and guide future tool design.

\section{Background}
In this section, we give a brief overview of Rust, 
its mechanism for concurrency safety,
and its thread communication and synchronization mechanisms.

\subsection{Rust and Rust-Based Software}
Rust is a low-level system programming language, firstly designed by Graydon Hoare in 2006. 
Rust has been sponsored and maintained by Mozilla since 2009 
and was firstly released in 2010~\cite{rust-wiki}.  
Providing safe concurrency~\cite{safe-concurrency-1,safe-concurrency-2}
and safe memory~\cite{safe-memory-1} with 
similar performance to C~\cite{rust-performance-1} is the design goal of Rust. 
In recent years, due to its safety and performance benefits, 
Rust draws many interests from developers. 
According to surveys on stack overflow, 
Rust is the most beloved programming language in the last 
three years~\cite{rust-love1,rust-love2,rust-love3}. 
Rust has become a popular language in building software systems, 
such as browsers~\cite{servo,quantum} and OSes~\cite{stratis,tockos,redox}.

\begin{table}[h!]
\centering
\footnotesize
{
 \setlength{\tabcolsep}{1mm}{
\begin{tabular}{|l|c|c|c|c|c|c|}
\hline
{\textbf{Application}}          &  {\textbf{Stars}} &  {\textbf{Commits}}  & {\textbf{Contributors}} & {\textbf{LOC}}   & {\textbf{ History}} \\ \hline \hline
Servo~\cite{servo}     						&    13150       &  35876     & 915    & 254K  			& 5.9 Years      \\ \hline
TiKV~\cite{tikv}    					    &    4381        &  3345      & 121    & 119K  			& 2.0 Years      \\ \hline
Rand~\cite{rand}                			&    359         &  1981      & 170    & 11K  			& 7.5 Years      \\ \hline

\end{tabular}}
}

  \mycaption{tab:apps}
{Information of selected applications.}
{The number of stars, commits, and contributors on GitHub, total source lines of code,
and development history on GitHub.}
\end{table}

In this paper, we select three popular open-source Rust projects 
on GitHub to conduct our study (Table~\ref{tab:apps}). 
According to the number of stars on GitHub, 
Servo is ranked 4th, TiKV is ranked 19th, 
and Rand is ranked in the top 3\% among all Rust projects. 
They cover different types of functionalities, 
including web browsing (Servo), 
key-value storage (TiKV), and library utility (Rand). 
They are of medium to large size 
and have at least two years' development history. 
We believe that these three applications are good 
representatives of real-world usage of Rust
and leave exploring other Rust applications to future work.

\subsection{Static Ownership Checking} 
\label{sec:checking}
Rust supports shared variables between threads. 
By default, one Rust value has exact one variable as its owner, and 
a value will be destroyed, if its owner variable leaves its scope. 
Rust's compiler conducts static ownership checking on shared variables, 
with the hope to reduce mistakes during accessing shared variables. 

First, the ownership of a value can be \textit{move}d from one scope to another, 
through function call or return, message passing, thread creation, and so on.
Rust compiler statically guarantees that an owner variable will not be
accessed after the ownership is moved to a different scope.
Therefore, a shared value can only be owned by one thread at any time. 

Second, Rust supports accessing a value using its reference. 
A reference can be \textit{borrow}ed from one scope to another, 
without moving ownership. 
Rust's compiler does not allow references to be borrowed between threads, 
since values' lifetime cannot be statically inferred across threads 
and Rust's compiler guarantees 
a value's lifetime covers all its reference usage. 

Third, to have multiple owner threads, 
Rust provides a thread-safe reference-counting pointer (\texttt{Arc}). 
Rust's compiler guarantees that all owner threads only access an \texttt{Arc} 
to read a shared variable or update a shared variable through \texttt{Arc::make\_mut()}, 
which will create a per-thread private copy of the shared variable.

Fourth, to allow multiple threads to update the same copy,
a shared variable has to be declared with both \texttt{Arc} and \texttt{Mutex}.
To access the shared variable, a thread needs to invoke 
the lock function of the shared variable's 
protecting \texttt{Mutex}.
A reference of the shared variable is returned
after successfully invoking the lock function, 
so that the thread can read or write the shared variable. 
Mutual exclusion provided by \texttt{Mutex} guarantees that 
there is at most one thread accessing a shared value at any time.

\subsection{Thread Synchronization in Rust}

\noindent{\underline{Thread Model.}}
Rust does not provide any runtime to manage thread scheduling. 
Rust threads are mapped to OS threads in an 1 to 1 way. 
To create a thread, \texttt{thread::spawn()} function is invoked 
with a piece of code (a \textit{closure}) as parameter. 
Ownership of shared variables can be moved from a creating 
thread to a created thread during thread creation.

\noindent{\underline{Thread Synchronization.}}
Similar to C/C++, Rust supports traditional synchronization 
primitives designed 
to protect shared memory accesses, 
such as lock (\texttt{Mutex}), read/write lock (\texttt{RwLock}), 
conditional variable (\texttt{Condvar}), atomic operations (\texttt{atomic}),
and barrier (\texttt{Barrier}). 
There are three difference between \texttt{Mutex} (or \texttt{RwLock}) 
in Rust and in C/C++.
First, Rust provides a poisoning mechanism~\cite{rust-poisoning} for 
\texttt{Mutex} to propagate panic information among threads.
Second, Rust requires a \texttt{Mutex} to be declared 
together with its protected data.
Third, there is no explicit unlock function in Rust, 
and Rust's compiler automatically adds unlock 
at the end of a scope with lock operations. 
Rust provides \texttt{Once} as a new primitive to 
guarantee the initializations of a global 
variable is only conducted once. 

Rust provides \texttt{channel} to pass messages between threads. 
Each channel has a sender and a receiver. 
Ownership can be moved from sender thread to receiver thread 
through message passing. 
When channel buffer is empty, 
pulling data from the channel will block.

\section{Usage Study}
\label{sec:usage}
In this section, we discuss our study results on
concurrency usage in Rust applications, including the usage of unsafe code 
and the usage of synchronization primitives.

\subsection{Unsafe Code}
Rust supports \texttt{unsafe} keyword to mark a function 
or a code scope to bypass some compiler checkings. 
As discussed in Section~\ref{sec:checking}, many static compiler 
checkings are designed to avoid concurrency bugs.
Therefore, the usage of \texttt{unsafe} 
is highly correlated with whether Rust can really achieve safe concurrency. 

\begin{table}[h!]
\centering
\scriptsize
{
\setlength{\tabcolsep}{0.1mm}{
\begin{tabular}{|l|c|c|c|c|c|c|c|}
\hline

\multirow{2}{*}{\textbf{Application}} & \multicolumn{4}{c|}{\textbf{\# of unsafe tags}} & \multicolumn{3}{c|}{\textbf{LOC}}  \\ \cline{2-8}
                                      & \textbf{fn} & \textbf{scope} & \textbf{total} & \textbf{\# per KLOC} & \textbf{average} & \textbf{percentage}   &  \textbf{total}                                  \\ \hline \hline

Servo                  & 31.72\% & 68.27\%    & 1683    &6.62   & 6.08    & 4.19\%             &  254K        \\ \hline
TiKV                   & 11.70\% & 88.29\%    & 94      & 0.78   & 5.62    & 0.44\%              & 119K          \\ \hline
Rand                   & 0       & 100.00\%   & 47      & 4.27   & 3.63    & 1.55\%            & 11K    \\ \hline
\end{tabular}
}
}
\mycaption{tab:unsafe}
{\texttt{unsafe} Usage.}
{
In the \# of unsafe tags columns, fn and scope: 
the ratio of \texttt{unsafe} used to tag a function or a code scope, 
and total: the total number of \texttt{unsafe}. 
In the LOC columns,
average: the average LOC tagged by an \texttt{unsafe}, 
and percentage: unsafe code over total code. 
}
\end{table}

We first study the amount of \texttt{unsafe} tags. 
As shown in Table~\ref{tab:unsafe},
there are a fair amount of \texttt{unsafe} tags 
in the three studied applications, 
ranging from 47 to 1683.  
We calculate the average number of \texttt{unsafe} 
tags over total source lines of code. 
On average, Servo developers use \texttt{unsafe} most frequently, 
and there are 6.62 \texttt{unsafe} tags per thousand lines of code. 
TiKV has the smallest number, 
and there is less than one \texttt{unsafe} tag 
per thousand lines of code. 
\texttt{unsafe} can be used to mark a function or a code scope. 
For the three studied applications, \texttt{unsafe} is constantly 
used more often to tag a code scope, 
ranging from 68.27\% to 100.00\% over all \texttt{unsafe} usage sites 
for the three applications.  
The large number of \texttt{unsafe} tags demonstrates
that Rust's compiler checkings
do restrict its programmability, 
encouraging programmers to bypass them using \texttt{unsafe}. 
These \texttt{unsafe} sites can potentially 
contain concurrency bugs, 
making Rust's concurrency less safe, 
as we will discuss in Section~\ref{sec:race}.

We then count the lines of code inside each code scope 
or function tagged with \texttt{unsafe}.
As shown in Table~\ref{tab:unsafe},
on average, an \texttt{unsafe} is used to tag a very small piece of code. 
For Servo, each \texttt{unsafe} tags 6.08 lines of code on average, 
which is the largest number among the three applications. 
The number drops to 5.62 and 3.63 for TiKV and Rand respectively. 
These results show that Rust developers are very careful 
when they use \texttt{unsafe}. 

In the end, we count the percentage of \texttt{unsafe} code 
over all code. 
As shown in Table~\ref{tab:unsafe}, the percentage of \texttt{unsafe} code 
ranges from 0.44\% to 4.19\% for the three studied applications. 
For Rust applications, safe code dominates.

\noindent{\underline{\bf{Observation 1:}}}
{\it{
The majority of code in Rust applications is safe. 
However, Rust developers often need to carefully use \texttt{unsafe} 
to bypass compiler checkings. 
}}

\noindent{\underline{\bf{Implication 1:}}}
{\it{
Compiler checkings do restrict Rust's programmability. 
Future works on more precise compiler checkings to give developers more flexibility
and verifying unsafe code in Rust to provide a certain 
level of confidence in safety are needed. 
}}

\subsection{Concurrency Usage}

Studying the usage of concurrency primitives can help understand 
how Rust developers implement thread communication and synchronization, 
and quantitatively get a feeling about which primitives are more likely to 
be misused, leading to concurrency bugs.

\begin{table}[h!]
\centering
\scriptsize
{
\setlength{\tabcolsep}{0.1mm}{
\begin{tabular}{|l|c|c|c|c|c|c|c|}
\hline

\multirow{2}{*}{\textbf{Application}} & \multicolumn{5}{c|}{\textbf{Shared Memory}}      & \multicolumn{1}{c|}{\textbf{Message Passing}}     & \multirow{2}{*}{\textbf{Total}} \\ \cline{2-7}
                                      & \textbf{Mutex} & \textbf{RwLock} & \textbf{Once} & \multicolumn{1}{c|}{\textbf{Condvar}} & \textbf{atomic}   &  \textbf{channel}  &                                 \\ \hline \hline

Servo                                & 14.42\%     & 12.16\%   & 0.21\%    & 0\%      & 11.66\%      & 61.52\%        &  1414        \\ \hline
TiKV                            & 8.40\%     & 12.36\%   & 0.09\%    & 18.69\%   & 18.69\%      & 42.43\%        & 1011          \\ \hline
Rand                                  & 6.25\%     & 15.62\%   & 9.37\%    & 0\%    & 62.5\%      & 6.25\%        & 32    \\ \hline
\end{tabular}
}
}
\mycaption{tab:usage}
{Concurrency Primitives Usage.}
{}
\end{table}

We count the number of operations for each primitive type. 
For example, for \texttt{Mutex}, we count the number of \texttt{Mutex.lock()} 
and \texttt{Mutex.try\_lock()} operations. 
As shown in Table~\ref{tab:usage}, \texttt{channel} is the most widely used 
primitive for two applications. 
For Servo, 62.52\% of synchronization operations are channel operations. 
For TiKV, the percentage is 42.53\%.
For Rand, its most widely used primitive is \texttt{atomic}, 
and the percentage is 
62.5\% over all primitive types. 
For multithreaded programs in other languages, 
mutex is the most widely used primitive~\cite{wangyin-1,wangyin-2,go-study-asplos}.
However, this is not true for Rust. 
We anticipate that it is Rust's compiler checkings that change 
developers' programming habits and usage patterns 
when building Rust concurrent software.

We also calculate the average number of concurrency operations
over lines of code. 
TiKV developers use concurrency operations most frequently,
and there are 8.49 concurrency operations for every thousand lines of code. 
Servo developers use concurrency operations less frequently,
and Rand developers use concurrency operations least frequently. 
The average operations per one thousand lines of code are 5.56 
and 2.90 for these two applications respectively.

\noindent{\underline{\bf{Observation 2:}}}
{\it{
The usage of concurrency primitive in Rust is different from other programming languages,
possibly caused by Rust's compiler checking mechanism.  
}}

\noindent{\underline{\bf{Implication 2:}}}
{\it{
Traditional concurrency bug detection and fixing techniques focus more on 
\texttt{Mutex}. Future research works should pay more attention to \texttt{channel} and \texttt{atomic}
to combat concurrency bugs in Rust. 
}}

\section{Concurrency Bug Study}
\label{sec:bug}

This section presents a preliminary study of Rust concurrency bugs. 
Specifically, we will first introduce our methodology of collecting and categorizing bugs
and then present our identified deadlock and data-race bugs.

\subsection{Methodology}

To collect concurrency bugs, we first search for the keywords ``\textit{deadlock}'' and ``\textit{race}'' in
GitHub commit histories of the three applications in Table~\ref{tab:apps}.
These two keywords are widely used in previous works to collect 
concurrency bugs in other programming languages~\cite{characteristics.asplos08,Hary-2,JaConTeBe,go-study-asplos}. 
Many previous works categorize 
concurrency bugs into deadlock bugs and non-deadlock bugs~\cite{characteristics.asplos08,Rui-FSE,hfix}.
Data race is one of the common types of non-deadlock bugs.

\begin{table}[h!]
\centering
\footnotesize
{
 \setlength{\tabcolsep}{0.3mm}{
\begin{tabular}{|l| >{\centering\arraybackslash}m{1.2cm}|>{\centering\arraybackslash}m{1.2cm}|>{\centering\arraybackslash}m{1.2cm}|>{\centering\arraybackslash}m{1.2cm}|}
\hline
 \multirow{2}{*}{\textbf{Application}}  & \multicolumn{2}{c|}{\textbf{Deadlock}}    & \multicolumn{2}{c|}{\textbf{Data Race}} \\ \cline{2-5}
                                        & \textbf{{Mutex}} & \textbf{{channel}} & \textbf{safe} & \textbf{unsafe} \\ \hline \hline
Servo 			                        & 5  & 3 & 1 & 5 \\ \hline
TiKV 		                            & 2  & 0 & 1 & 0 \\ \hline
Rand 			                        & 0  & 0 & 1 & 0 \\ \hline
\textbf{Total} 	                        & 7  & 3 & 3 & 5 \\ \hline

\end{tabular}
}
\mycaption{tab:bug}
{Taxonomy.}
{This table shows how our studied bugs distribute across different categories and applications.}
}
\end{table}

We then manually inspect the resulting commits that contain the two keywords
to identify real concurrency bugs.
In the last step,
we study identified concurrency bugs,
by referring their patches and related discussions on GitHub. 
Right now, we have studied 18 bugs.
Their detailed 
distribution is shown in Table~\ref{tab:bug}.

\subsection{Rust Deadlock Bugs}

In total, we study 10 deadlock bugs. 
We categorize them based on which synchronization primitives causing 
threads not to make future progress. 
As shown in Table~\ref{tab:bug}, 
due to implicit unlock in Rust, double locks are the root causes for seven deadlocks.
The other three deadlocks are caused by misusing channel.

{\vspace{-0.12in}
\begin{figure}[th]
\begin{center}
\scriptsize
\begin{lstlisting}[xleftmargin=.15in,language=Rust,basicstyle=\ttfamily,morekeywords={-},morekeywords={+},keepspaces=true]
 pub fn insert_rule(...) -> Fallible<u32> {
-  let mut guard = stylesheet.shared_lock.write();
-  let new_rule = rules.write_with(&mut guard)?;
+  let new_rule = {
+    let mut guard = stylesheet.shared_lock.write();
+    rules.write_with(&mut guard)?
+  }
   ...
   let dom_rul = CSSRule::new_specific(...);
 }

 pub fn new_specific(...) -> Root<CSSRule> {
   ...
   let guard = stylesheet.shared_lock().read();
 } 
\end{lstlisting}
\vspace{-0.15in}
  \mycaption{fig:double-lock}
{A deadlock caused by double \texttt{RwLock}.}
{The code has been simplified to ease our explannation.}

\end{center}
\vspace{-0.15in}
\end{figure}
}

\noindent{\underline{Implicit Unlock.}}
Rust does not provide any unlock function.
Rust's compiler automatically adds corresponding unlock functions 
at the end of a scope with lock functions.
Due to this, Rust developers may forget to 
release acquired lock timely, 
leading to double-lock deadlocks. 

One bug example from Servo is shown in Figure~\ref{fig:double-lock}. 
\texttt{sheet.shared\_lock} is a \texttt{RwLock}.
The write lock is acquired at line 2. 
The write lock is only needed for the invocation of 
function \texttt{write\_with()}, 
but it is not released until the end 
of function \texttt{insert\_rule()} at line 10. 
Unfortunately, \texttt{insert\_rule()} invokes \texttt{new\_specific()} 
at line 9. \texttt{new\_specific()} acquires 
the read lock of the \texttt{RwLock} at line 14, 
leading to a double-lock bug. 
To fix this bug, Servo developers simply 
add a pair of \texttt{\{\}} to create a new scope, 
so that the write lock can be released after 
invoking \texttt{write\_with()} at line 7.

\noindent{\underline{Misusing Channel.}}
As shown in Table~\ref{tab:usage}, 
channel is widely used in Rust programs. 
When buffer is empty, pulling data from a channel will block,
forming an edge from receiver thread 
to sender thread in wait-for graph.
A circle in wait-for graph means circular wait among threads and deadlock. 
We have three bugs caused by circular wait involving waiting for 
channel message.

\begin{figure}
\centering
\subfloat[master thread]{
  \scriptsize
\begin{lstlisting}[xleftmargin=.05in,language=Rust,basicstyle=\ttfamily,morekeywords={-},morekeywords={+},keepspaces=true]
while !done {
 match r1.recv(){
  Some(Exit(...)) => {
   done = true;
   ...
  }
  Some(SetID(s2)) => {
   s2.send(());
  }
  ...
 }
}

let (s3, r3) = channel();
s2.send(ExitMsg(s3));
r3.recv(); //blocks
\end{lstlisting}
  \label{fig:master}
  } 
  \hspace{2mm}‎
\subfloat[worker thread]{
 \scriptsize
\begin{lstlisting}[xleftmargin=.05in,language=Rust,basicstyle=\ttfamily,morekeywords={-},morekeywords={+},keepspaces=true]
loop {
 let request = r2.recv();
  match request {
   ExitMsg(s3) => {
    ...
    s3.send(());
    break;
   }
   ReadyMsg(...) => {
    ...
    let (s2, r2) = channel();
    s1.send(SetID(s2))
    r2.recv() //blocks
   }
 }
}
\end{lstlisting}
 \label{fig:worker}
 }
\mycaption{fig:channel}{A deadlock caused by circularly waiting for message between two threads.}
{s: sender; r: receiver; numbers after s or r are used to match sender and receiver.}
\label{fig:time} 
\end{figure} 

One example from Servo is shown in Figure~\ref{fig:channel}. 
There are two threads, master thread and worker thread.
Both of them have a message-processing loop to 
handle messages from \texttt{r1} and \texttt{r2} respectively. 
In Figure~\ref{fig:worker}, 
when work thread wants to set its ID, 
it creates a channel at line 11, send the sender \texttt{s2} 
of the created channel in a \texttt{SetID} message to master thread 
at line 12, 
and blocks itself to wait for reply from master thread at line 13. 
In Figure~\ref{fig:master}, 
when master thread receives a \texttt{SetID} message from \texttt{r1}, 
it replies work thread by sending an empty 
message \texttt{()} using \texttt{s2} to unblock worker thread. 
When master thread receives an \texttt{Exit} message at line 3, 
it leaves the message-handling loop at line 4.
After that, master thread notifies worker thread to exit at line 15,
and blocks itself to wait for a reply from worker thread at line 16. 
Worker thread can unblock master thread by sending out an empty 
message \texttt{()} at line 6 in Figure~\ref{fig:worker}.

This bug happens when master thread receives an \texttt{Exit} message just 
before worker thread sends out a \texttt{SetID} message. 
Master thread blocks at line 16 in Figure~\ref{fig:master} 
waiting for a message should be sent from work 
thread at line 6 in Figure~\ref{fig:worker}.
However, worker thread blocks at line 13 in Figure~\ref{fig:worker}, 
waiting for master thread to unblock it by execution 
line 8 in Figure~\ref{fig:master}. 
To fix this bug, Servo developers change master thread, 
and only allow it to leave the message-processing loop 
after worker thread has exit,
so that master thread can always handle messages from worker thread.

\noindent{\underline{\bf{Observation 3:}}}
{\it{
Double locks and misusing channels are two common causes of deadlocks in Rust. 
}}

\noindent{\underline{\bf{Implication 3:}}}
{\it{
To detect deadlocks in Rust, it is useful to design static analysis which can 
identify possible lock operations in a function call and 
can infer thread wait relationship based on inter-thread messaging.
}}

\subsection{Rust Data Races}
\label{sec:race}
Intuitively, there could be data races inside unsafe code. 
Our study confirms this intuition.
In total, five out of eight races are caused by 
instructions inside unsafe code. 
However, what is counter-intuitive is that there are still races inside safe code.

\noindent{\underline{Data Races in Safe Code.}}
Rust supports shared variables in atomic types, 
such as \texttt{AtomicBool}, \texttt{AtomicPtr}, and \texttt{AtomicUsize}.
Read and write conducted on atomic variables are automatically ignored 
by Rust's ownership checkings. 
All races in safe code are caused by misusing atomic operations.

{
\begin{figure}[th]
\begin{center}
\scriptsize
\begin{lstlisting}[xleftmargin=.15in,language=Rust,basicstyle=\ttfamily,morekeywords={-},morekeywords={+},keepspaces=true]
-  static CHECKED: AtomicBool = ATOMIC_BOOL_INIT;
+  static CHECKER: Once = ONCE_INIT;
   static AVAILABLE: AtomicBool = ATOMIC_BOOL_INIT;

   fn is_getrand_available() -> bool {
-    if !CHECKED.load(Ordering::Relaxed) {
+    CHERKER.call_once(|| {
       let mut buf: [u8; 0] = [];
       let result = getrand(&mut buf);
       let available = if result == -1 {
         ...
       } else {
         true
       };
       AVAILABLE.store(available, Ordering::Relaxed);
-      CHECKED.store(true, Ordering::Relaxed);
-      available
-    } else {
-      AVAILABLE.load(Ordering::Relaxed)
     }
     AVAILABLE.load(Ordering::Relaxed)
   }
\end{lstlisting}
\caption{An atomicity violation bug in safe code.}
\label{fig:atomicity}
\end{center}
\end{figure}
}

An example from Rand is shown in Figure~\ref{fig:atomicity}. 
Function \texttt{is\_getrand\_available()} invokes \texttt{getrand()} at line 7,
and checks whether the returned value is \texttt{-1} 
at line 8 to decide whether \texttt{getrand()} is available.
\texttt{getrand()} is an expensive system call,  
so that Rand developers decide to cache the previous execution result.
Since  \texttt{is\_getrand\_available()}  can be executed concurrently,
two atomic variables, \texttt{CHECKED} and \texttt{AVAILABLE}, 
are introduced to achieve the caching functionality.
\texttt{CHECKED} represents whether \texttt{getrand()} is called before, 
and \texttt{AVAILABLE} represents whether \texttt{getrand()} is available. 
If \texttt{CHECKED} is \texttt{false} at line 5, 
\texttt{getrand()} will be called at line 7, 
\texttt{AVAILABLE} will be set to suitable value at line 13, 
and \texttt{CHECKED} will be set to \texttt{true} at line 14.
If \texttt{CHECKED} is \texttt{true} at line 5, 
the value of \texttt{AVAILABLE} will be returned at line 19.

There is a concurrency bug that can cause  
\texttt{is\_getrand\_available()} to return incorrect results. 
Since either compiler or CPU could reorder the execution of line 13 and line 14,
if the reordering happens, 
the caching mechanism is only correct 
when line 14 and line 13 are executed atomically.  
If the execution of line 14 and line 13 of one thread 
is interleaved by the execution of line 5 and line 19 of another thread, 
\texttt{is\_getrand\_available()} will return \texttt{false}, 
no matter \texttt{getrand()} is available or not. 
To fix this bug, Rand developers use \texttt{Once} primitive to guarantee 
the invocation of \texttt{getrand()} and the initialization of \texttt{AVAILABLE}
are only conducted once and are conducted atomically.

\noindent{\underline{\bf{Observation 4:}}}
{\it{
Unsafe code is indeed one major source of data races in Rust. 
However, there are also data races from safe code. 
}}

\noindent{\underline{\bf{Implication 4:}}}
{\it{
Race detection techniques are needed for Rust, and they should 
focus on unsafe code and atomic operations in safe code. 
}}

\section{Related Works}
\label{sec:related}
Many system reliability researchers conducted 
empirical studies on real-world bugs before~\cite{lift-study,chou01empirical,
characteristics.asplos08,Lu.study.fast,sullivan92comparison,Rui-FSE,PerfStudy,Hary-1,Hary-2,go-study-asplos,junwen-1}.
These works have successfully guided techniques 
to combat bugs from various aspects.
To the best of our knowledge, our work is the first empirical 
study on Rust concurrency usage and concurrency bugs.
Although it is still at an initial stage,
our current results can inspire and guide future
techniques for concurrency bugs in Rust. 

\section{Conclusion and Future Work}
With the increasing usage of Rust to implement various concurrent systems,
it is important to understand whether Rust's static compiler 
checkings can really bring safe concurrency. 
We conduct the first study on Rust concurrency from two aspects: 
concurrency usage and real-world  concurrency bugs.
We expect our study to deepen the understanding of Rust's concurrency 
and motivate more research works on Rust. 

This paper is just a starting point to understand Rust.
We plan to extend the current work from the following directions. 

First, more concurrency bugs should be collected from a wider range of Rust applications.
We should build a more comprehensive taxonomy for Rust concurrency bugs 
from various aspects, such as root causes, fix strategies, how bugs are introduced, and so on. 

Second, we plan to systematically evaluate existing concurrency bug 
detection techniques on Rust. Rust uses LLVM as its backend and doesn't have any runtime. 
It is fairly easy to apply existing techniques designed for C/C++ to Rust applications.  
Since Rust has message-passing mechanisms, existing techniques are expected 
to be extended to have more precise detection results.

Third, safe memory usage is another design goal of Rust. 
We plan to extend our study to memory bugs in Rust and 
understand whether Rust's compiler checkings can 
really reduce memory bugs. 
Many memory bugs can be exploited by hackers have large security impact. 
We also plan to evaluate security impact of memory bugs in Rust. 

Four, Rust is a very young language, and new 
language features were added in recent years. 
We plan to study new language 
features influence Rust's programmability and its safety.


\balance
{
\bibliographystyle{abbrvnat}
\bibliography{main} 
}
\end{document}